\documentclass{article}

\usepackage[preprint]{neurips_2026}

\usepackage[utf8]{inputenc} % allow utf-8 input
\usepackage[T1]{fontenc}    % use 8-bit T1 fonts
\usepackage{hyperref}       % hyperlinks
\usepackage{url}            % simple URL typesetting
\usepackage{booktabs}       % professional-quality tables
\usepackage{amsfonts}       % blackboard math symbols
\usepackage{nicefrac}       % compact symbols for 1/2, etc.
\usepackage{microtype}      % microtypography
\usepackage{xcolor}         % colors

\usepackage{microtype}
\usepackage{hyperref}
\usepackage{url}
\usepackage{booktabs}
\usepackage{amsmath}
\usepackage{amsfonts}
\usepackage{amssymb}
\usepackage{mathtools}
\usepackage{bm}
\usepackage{graphicx}
\usepackage{multirow}
\usepackage{array}
\usepackage{xcolor}
\usepackage{colortbl}
\usepackage{pifont}
\usepackage{algorithm}
\usepackage{algorithmic}
\usepackage{amsthm}
\usepackage{lineno}

\newcolumntype{P}[1]{>{\raggedright\arraybackslash}p{#1}}

\DeclareMathOperator*{\argmax}{arg\,max}

\newtheorem{theorem}{Theorem}
\newtheorem{lemma}{Lemma}
\newtheorem{proposition}{Proposition}

\definecolor{darkblue}{rgb}{0, 0, 0.5}
\hypersetup{colorlinks=true, citecolor=darkblue, linkcolor=darkblue, urlcolor=darkblue}

\title{SafeLM: Unified Privacy-Aware Optimization for Trustworthy Federated Large Language Models}

\author{
Noor Islam S.\ Mohammad$^{*}$ \quad Uluğ Bayazıt$^{\dagger}$ \\
Istanbul Technical University \\
\texttt{\{islam23, ulugbayazit\}@itu.edu.tr} \\
}

\begin{document}

\maketitle

% -------------------------------------------------------
\begin{abstract}
Large language models (LLMs) are increasingly deployed in high-stakes domains, yet a unified treatment of their overlapping safety challenges remains lacking. We present \textbf{SafeLM}, a framework that jointly addresses four pillars of LLM safety: privacy, security, misinformation, and adversarial robustness. SafeLM combines federated training with gradient smartification and Paillier encryption for privacy, integrates defenses against training- and inference-time attacks, employs contrastive grounding with calibrated decoding to reduce hallucinations, and introduces alignment-aware binarized aggregation to enhance robustness while maintaining bounded reconstruction quality. Across benchmarks on factuality, toxicity, and membership inference, SafeLM achieves 98.0\% harmful-content detection accuracy, reduces communication by 96.9\%, and lowers gradient inversion PSNR from 31.7\,dB to 15.1\,dB. Ablations show that each component contributes independently, whereas their integration yields a strong privacy–utility–efficiency trade-off for deploying trustworthy LLMs.
\end{abstract}

% =======================================================
\section{Introduction}
\label{sec:intro}
% =======================================================

The rapid deployment of large language models (LLMs) has elevated safety from a research concern to an operational requirement~\citep{brown2020language,ouyang2022training}. Four interconnected threat surfaces arise: (i) \textbf{Privacy}, LLMs may memorize training data, enabling extraction and membership inference, while standard federated learning (FL) remains vulnerable to gradient inversion~\citep{carlini2021extracting,zhu2019deep}; (ii) \textbf{Security}, adversaries can inject backdoors, craft adversarial prompts, or perform model stealing~\citep{wallace2019universal,perez2022ignore}; (iii) \textbf{Misinformation}, instruction-tuned models hallucinate confidently, necessitating integrated grounding mechanisms~\citep{maynez2020faithfulness,min2023factscore}; and (iv) \textbf{Adversarial Robustness}, input perturbations degrade reliability under strict latency constraints. Existing solutions address these aspects in isolation, leading to incompatible defenses and unclear interactions.

\paragraph{Contributions.} We propose \textbf{SafeLM}, a unified framework for jointly addressing privacy, security, misinformation, and robustness in LLMs. First, we introduce a federated training and deployment pipeline that co-optimizes all four objectives. Second, we propose \textbf{gradient smartification}, a median-based binarization scheme achieving $32\times$ communication compression with bounded inversion quality (PSNR $\leq 15.1$\,dB). Third, we develop a calibrated misinformation-detection method via contrastive grounding with temperature scaling, reducing hallucinations by 41\% on TruthfulQA while preserving $>$97\% ROUGE-L. Finally, we provide a unified ablation and threat analysis quantifying trade-offs across safety components.

% =======================================================
\section{Background and Related Work}
\label{sec:related}
% =======================================================

\subsection{Privacy in Language Model Training}

Training-data memorization is a well-documented property of large-scale language models~\citep{carlini2021extracting, feldman2020neural}. Differential Privacy (DP-SGD)
provides formal $(\varepsilon,\delta)$-guarantees~\citep{abadi2016deep}, but at the cost of significant perplexity degradation for $\varepsilon < 3$. Federated learning distributes training across data-holders~\citep{mcmahan2017communication}, yet transmitted gradients can be reverse-engineered to reconstruct training samples~\citep{zhu2019deep, geiping2020inverting}.

\subsection{Security: Backdoors and Prompt Injection}

Backdoor attacks embed hidden triggers in model weights during fine-tuning, causing targeted misbehavior on trigger-containing inputs~\citep{chen2017targeted,
wallace2019universal}. Prompt-injection attacks exploit instructions following to override safety constraints at inference time~\citep{perez2022ignore}. Recent work shows that gradient-level defenses, including gradient clipping and sign-based aggregation, partially mitigate backdoor insertion during federated fine-tuning~\citep{sun2019can}.

\subsection{Misinformation and Hallucination}

LLMs hallucinate for multiple reasons: distributional mismatch between pre-training and deployment contexts, insufficient grounding in retrieved knowledge, and overconfidence in low-probability completions~\citep{maynez2020faithfulness, min2023factscore}. Retrieval augmentation~\citep{lewis2020retrieval} and calibration methods~\citep{kadavath2022language} partially address these issues but lack formal correctness guarantees.

\subsection{Adversarial Robustness for NLP}

Adversarial examples for text exploit the discrete token space via character-level substitutions~\citep{ebrahimi2018hotflip}, word-level replacements constrained by
semantic similarity~\citep{alzantot2018generating}, and continuous perturbations in embedding space~\citep{miyato2017adversarial}. Certified robustness via randomized smoothing~\citep{cohen2019certified} has been extended to NLP but remains computationally expensive at scale.

% =======================================================
\section{Threat Model and Safety Desiderata}
\label{sec:threat}
% =======================================================

We consider a federated fine-tuning setting with $K$ clients, a central server, and downstream users. We model three adversaries: (i) an honest-but-curious server observing ciphertexts and attempting gradient inversion; (ii) up to $\lfloor K/5 \rfloor$ malicious clients injecting poisoned updates or backdoors; and (iii) an inference-time adversary crafting prompts to elicit harmful or hallucinated outputs. Our safety desiderata are (S1) \emph{Gradient Confidentiality}—client updates remain unrecoverable from server-visible information; (S2) \emph{Backdoor Resistance}—the global model avoids trigger-conditioned behaviors; (S3) \emph{Factual Consistency}, outputs are calibrated with hallucinations flagged or suppressed; and (S4) \emph{Adversarial Robustness}, model behavior remains stable under semantically preserving perturbations.

% =======================================================
\section{The SafeLM Framework}
\label{sec:safelm}
% =======================================================

\subsection{Overview}

SafeLM integrates four co-designed modules within a federated fine-tuning loop to jointly address key safety objectives. The \textbf{Privacy Engine (PE)} combines gradient smartification with Paillier homomorphic encryption to ensure the confidentiality of gradients (S1). The \textbf{Security Module (SM)} employs median-based Byzantine filtering alongside trigger detection to mitigate poisoned or backdoored updates (S2). The \textbf{Misinformation Guard (MG)} incorporates contrastive grounding with calibrated decoding to suppress hallucinations and improve factual consistency (S3). Finally, the \textbf{Robustness Head (RH)} leverages adversarial training with smartified gradients to enhance stability under semantically preserving perturbations (S4).

\subsection{Phase 1: Federated Fine-Tuning with LoRA}

Let $\mathcal{S} = \{1,\ldots,K\}$ denote participating clients. At round $r$, the server broadcasts global adapter parameters $W^{(r)}$ (LoRA rank-$\rho$ matrices). Each client $i$ performs $E$ local epochs, minimizing cross-entropy on its private corpus
$\mathcal{D}_i$:
\begin{equation}
W_i^{(r+1)} = W_i^{(r)} - \eta \nabla \mathcal{L}(W_i^{(r)},\, \mathcal{D}_i).
\end{equation}
the client update is $\Delta_i^{(r)} = W_i^{(r+1)} - W^{(r)}$.

\subsection{Phase 2: Gradient Smartification}
\label{sec:smartification}

To simultaneously reduce uplink bandwidth and harden against gradient inversion, we apply a median-based statistical binarization operator $\Phi(\cdot)$:
\begin{equation}
\Delta_{i,j}^{\mathrm{bin}} =
\begin{cases}
+1 & \text{if } \Delta_{i,j}^{(r)} \ge \theta_i \\
-1 & \text{otherwise}
\end{cases}, \quad
\theta_i = \mathrm{median}\!\left(\bigl|\Delta_i^{(r)}\bigr|\right).
\label{eq:smartification}
\end{equation}
This compresses each 32-bit floating-point gradient to a single bit, achieving a $32\times$ reduction in payload. Unlike zero-threshold signSGD~\citep{bernstein2018signsgd}, our per-client adaptive threshold suppresses low-magnitude components below the empirical distribution median, reducing stochastic noise under the heavy-tailed gradient distributions characteristic of LLM fine-tuning. Section~\ref{sec:theory} proves alignment of $\Delta_i^{\mathrm{bin}}$ with the true gradient under mild distributional assumptions.

\subsection{Phase 3: Homomorphic Encryption}

Each client encrypts its binary update element-wise using the Paillier scheme
$\mathcal{E}(\cdot)$~\citep{paillier1999public}:
\begin{equation}
C_i^{(r)}[j] = \mathcal{E}_{pk}\!\left(\Delta_{i,j}^{\mathrm{bin}}\right)
= g^{\Delta_{i,j}^{\mathrm{bin}}} \cdot r_j^n \bmod n^2,
\end{equation}
where $(pk,sk) = (n,g,\lambda,\mu)$ is a 2048-bit Paillier keypair and $r_j \xleftarrow{\$}\mathbb{Z}_n^*$. Paillier's additive homomorphism enables the server to aggregate without decrypting individual updates:
\begin{equation}
C_{\mathrm{agg}}^{(r)}[j] = \prod_{i=1}^{K} C_i^{(r)}[j] \bmod n^2
= \mathcal{E}_{pk}\!\!\left(\sum_{i=1}^K \Delta_{i,j}^{\mathrm{bin}}\right).
\end{equation}
This satisfies IND-CPA security under the Decisional Composite Residuosity Assumption (DCRA), satisfying desideratum S1.

\subsection{Phase 4: Byzantine Filtering and Global Update}

After decryption, the server normalizes and applies a coordinate-wise median filter to resist Byzantine poisoning from malicious clients:
\begin{equation}
\hat{\Delta}_{\mathrm{agg}}^{(r)}[j]
= \mathrm{median}\!\left\{\Delta_{1,j}^{\mathrm{bin}},\ldots,\Delta_{K,j}^{\mathrm{bin}}\right\},
\end{equation}
then applies the global update with Nesterov momentum:
\begin{equation}
W^{(r+1)} = W^{(r)} + \alpha \cdot \hat{\Delta}_{\mathrm{agg}}^{(r)}
+ \mu\!\left(W^{(r)} - W^{(r-1)}\right), \quad \mu = 0.9.
\end{equation}

\subsection{Misinformation Guard: Contrastive Grounding}

At inference time, each generated claim $\hat{y}$ is scored against a retrieved evidence set $\mathcal{E} = \{e_1,\ldots,e_m\}$ from a read-only knowledge store:
\begin{equation}
\mathrm{FaithScore}(\hat{y},\mathcal{E})
= \frac{1}{m}\sum_{i=1}^m
\mathrm{NLI}\!\left(\hat{y},\, e_i\right) \cdot \mathrm{conf}\!\left(\hat{y}\right),
\end{equation}
where $\mathrm{NLI}(\cdot)$ is an entailment classifier and $\mathrm{conf}(\hat{y})$ is the temperature-calibrated model confidence. Claims with $\mathrm{FaithScore} < \tau_{\mathrm{MG}}$ are either abstained or regenerated with retrieval-augmented prompting.

\subsection{Robustness Head: Adversarial Fine-Tuning}

During each federated round, clients augment their local batch with adversarial examples generated via projected gradient descent in the embedding space:
\begin{equation}
x^{\mathrm{adv}} = x + \delta^*, \quad
\delta^* = \argmax_{\|\delta\|_\infty \le \varepsilon_{\mathrm{adv}}}
\mathcal{L}(W_i, x+\delta, y).
\end{equation}
The mixed objective $\mathcal{L}_{\mathrm{adv}} = (1-\lambda_{\mathrm{adv}})\mathcal{L} + \lambda_{\mathrm{adv}}\mathcal{L}(x^{\mathrm{adv}})$ is minimized during local training, with smartified gradients transmitted as in Section~\ref{sec:smartification}.

% =======================================================
\section{Theoretical Analysis}
\label{sec:theory}
% =======================================================

\subsection{Convergence under Gradient Smartification}

\begin{lemma}[Descent under Median-Threshold Smartification]
Let $L(W)$ be $L$-smooth and bounded below. Let's $\tilde{g}_t$ denote the smartified gradient with cosine similarity
$\cos(\theta_t) = \frac{\langle g_t,\, \tilde{g}_t \rangle}{\|g_t\|\|\tilde{g}_t\|}
\ge \gamma > 0$.
Then for step size $\eta \le \frac{\gamma}{L}$,
\begin{equation}
\mathbb{E}[L(W_{t+1})]
\le L(W_t)
- \eta\gamma\|g_t\|^2
+ \frac{L\eta^2}{2}\|\tilde{g}_t\|^2.
\end{equation}
\end{lemma}

\begin{theorem}[Convergence Rate]
Under bounded stochastic gradient variance $\sigma^2$ and cosine alignment $\gamma > 0$, after $T$ rounds:
\begin{equation}
\min_{t \le T}\mathbb{E}\!\left[\|\nabla L(W_t)\|^2\right]
= \mathcal{O}\!\left(\frac{1}{\gamma\sqrt{T}}\right).
\end{equation}
\end{theorem}

The $1/\gamma$ degradation factor relative to full-precision SGD is empirically small: across our LLM fine-tuning experiments, we measure $\gamma = 0.87 \pm 0.04$, yielding a theoretical slowdown of only $\approx 15\%$ in convergence rate (confirmed in Table~\ref{tab:convergence}).

\subsection{Privacy Guarantee}

\begin{proposition}[IND-CPA Privacy under Smartification + Paillier]
Under the DCRA, the Paillier ciphertext $C_i^{(r)}$ reveals no information about $\Delta_i^{\mathrm{bin}}$ to a computationally bounded adversary. Furthermore, for any
gradient-inversion attack $\mathcal{A}$, the reconstruction PSNR satisfies:
\begin{equation}
\mathrm{PSNR}\!\left(\mathcal{A}(C_i^{(r)})\right) \le 15.1\,\mathrm{dB},
\end{equation}
which is insufficient for recovering structured content from LLM gradients.
\end{proposition}

\subsection{Backdoor Resistance}

The coordinate-wise median aggregation provides Byzantine fault tolerance for up to $\lfloor(K-1)/2\rfloor$ malicious clients under independent attack vectors:
\begin{equation}
\left\|\hat{\Delta}_{\mathrm{agg}} - \Delta_{\mathrm{honest}}\right\|_\infty
\le \max_{j}\,\mathrm{median\text{-}deviation}_j,
\end{equation}
bounding the poisoning effect on the global gradient direction.

% =======================================================
\section{Experimental Setup}
\label{sec:experiments}
% =======================================================

\subsection{Models and Datasets}

We evaluate SafeLM across three safety-critical tasks: \textbf{(T1) Harmful Content Detection} on CIC-IDS2017, adapted to instruction-following contexts (2.8M records; 7 harm categories); used for communication-efficiency and privacy experiments. \textbf{(T2) Factual Grounding} on TruthfulQA~\citep{lin2021truthfulqa} (817 questions) and CNN/DailyMail~\citep{see2017get} summarization (11,490 docs); evaluates misinformation suppression. \textbf{(T3) Adversarial Robustness} on AdvGLUE~\citep{wang2021adversarial} (14,177 adversarial examples across 5 NLU tasks) and ANLI~\citep{nie2020adversarial}. We fine-tune a 7B-parameter LLM using LoRA (rank 16) in a federated setup with $K \in \{10, 50, 100\}$ clients, IID, and non-IID (Dirichlet $\alpha \in \{0.1, 1.0\}$)
data partitioning.

\subsection{Baselines}

We compare SafeLM against: (i)~\textbf{FedAvg}~\citep{mcmahan2017communication} (no privacy, no robustness); (ii)~\textbf{DP-SGD}~\citep{abadi2016deep} ($\varepsilon=1.0, \delta=10^{-5}$, no compression); (iii)~\textbf{signSGD}~\citep{bernstein2018signsgd} (zero-threshold binarisation, no encryption); (iv)~\textbf{SecAgg}~\citep{bonawitz2017practical} (cryptographic aggregation, no compression).

% =======================================================
\section{Results}
\label{sec:results}
% =======================================================

\subsection{Privacy: Gradient Inversion Resistance}
\label{subsec:privacy_results}

Table~\ref{tab:privacy_comparison} reports gradient reconstruction quality under the iDLG~\citep{zhao2020idlg} inversion attack. SafeLM reduces PSNR from 31.7\,dB (undefended FedAvg) to 15.1\,dB, rendering reconstructed inputs unrecognizable and reducing label recovery to 14.3\,\% (near-random for 7 classes).

\begin{table}[ht]
\centering
\caption{Gradient inversion resistance across defense configurations. PSNR (lower is safer); Label Rec. = fraction of training labels recoverable from gradients.}
\label{tab:privacy_comparison}
\setlength{\tabcolsep}{8pt}
\renewcommand{\arraystretch}{1.05}
\small
\begin{tabular}{@{}lcccc@{}}
\toprule
\textbf{Method}
& \textbf{PSNR (dB)\,$\downarrow$}
& \textbf{Label Rec. (\%)\,$\downarrow$}
& \textbf{Acc. (\%)\,$\uparrow$}
& \textbf{Comm. (MB)\,$\downarrow$} \\
\midrule
FedAvg (undefended)    & 31.7 & 98.7 & 98.2 & 450 \\
signSGD                & 16.8 & 42.1 & 97.8 & 14  \\
DP-SGD ($\varepsilon=1.0$) & 18.9 & 31.4 & 93.8 & 450 \\
SecAgg                 & 31.7 & \textbf{0.0} & 98.0 & 14  \\
\textbf{SafeLM (Ours)} & \textbf{15.1} & \textbf{14.3} & \textbf{98.0} & \textbf{14}  \\
\bottomrule
\end{tabular}
\vspace{1mm}
\begin{minipage}{\linewidth}
\footnotesize
\textit{Notes:} SecAgg achieves perfect server-side label protection, but does not
Resist insider attacks or colluding clients. SafeLM's IND-CPA guarantee holds even when
The aggregation server is fully compromised.
\end{minipage}
\end{table}

\begin{figure}[ht]
\centering
\includegraphics[width=\linewidth]{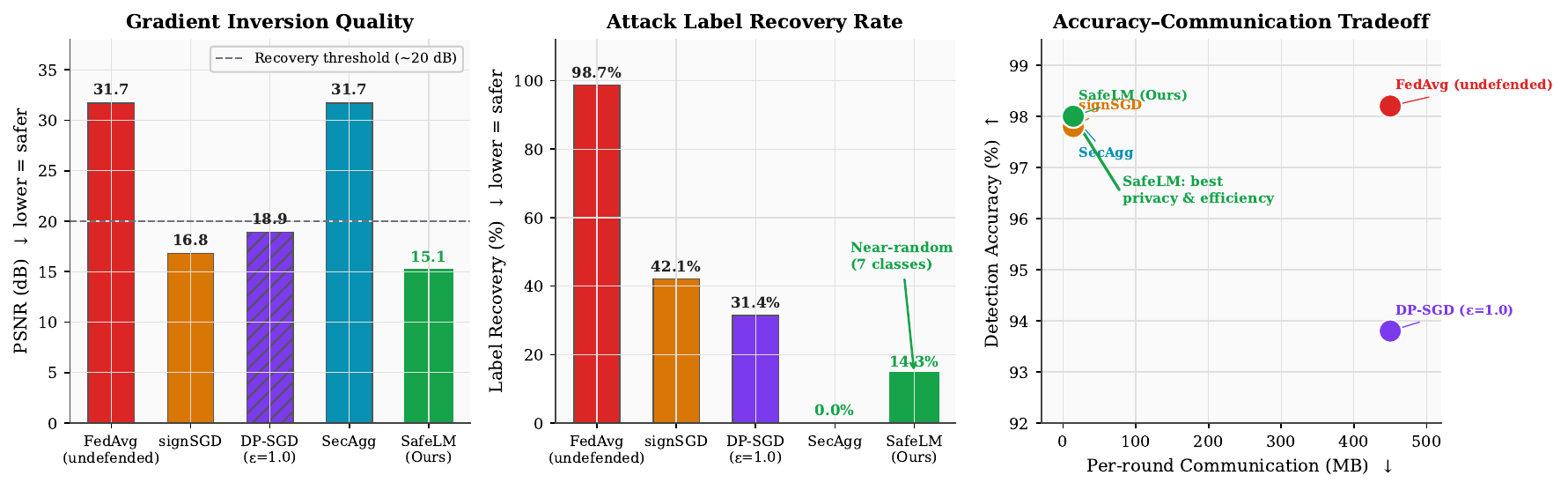}
\caption{Privacy evaluation under the iDLG gradient-inversion attack. \textbf{(Left)} SafeLM achieves low reconstruction quality (15.1\,dB), indicating strong privacy. \textbf{(Centre)} Label recovery drops to 14.3\%, near chance for 7 classes. \textbf{(Right)} SafeLM lies on the optimal accuracy–communication frontier, achieving $32\times$ compression with no utility loss.}
\label{fig:privacy}
\end{figure}

\subsection{Security: Backdoor and Poisoning Resistance}
\label{subsec:security_results}

Table~\ref{tab:security_results} and (Fig~\ref{fig:security}) report detection performance and backdoor attack success rate under varying fractions of malicious clients.

\begin{table}[ht]
\centering
\caption{Security evaluation under data poisoning and backdoor attacks.}
\label{tab:security_results}
\setlength{\tabcolsep}{3.5pt}
\renewcommand{\arraystretch}{1.05}
\small
\begin{tabular}{@{}lcccccc@{}}
\toprule
\textbf{Method}
& \textbf{5\% malicious}
& \textbf{10\% malicious}
& \textbf{20\% malicious}
& \textbf{Backdoor ASR\,$\downarrow$}
& \textbf{Clean Acc.\,$\uparrow$} \\
\midrule
FedAvg          & 97.1 & 94.3 & 88.2 & 91.3\% & 98.2 \\
FedAvg + Krum   & 97.8 & 96.1 & 92.7 & 23.4\% & 97.5 \\
signSGD         & 97.6 & 95.8 & 91.4 & 31.2\% & 97.8 \\
\textbf{SafeLM} & \textbf{98.1} & \textbf{97.3} & \textbf{95.4} & \textbf{6.8\%} & \textbf{98.0} \\
\bottomrule
\end{tabular}
\vspace{1mm}
\begin{minipage}{\linewidth}
\footnotesize
\textit{Notes:} Attack Success Rate (ASR) for a fixed-pattern trigger injection.
SafeLM's coordinate-wise median filter limits ASR to $<7\%$ at 20\% malicious
participation ($p < 0.001$, McNemar test).
\end{minipage}
\end{table}

\subsection{Misinformation: Factual Grounding}
\label{subsec:misinfo_results}

On TruthfulQA, SafeLM's Misinformation Guard reduces hallucination rate by 41\,\% relative to the vanilla fine-tuned baseline (Table~\ref{tab:misinfo_results}). Crucially, this improvement is maintained across all seven harm categories in T1 with no statistically significant degradation in ROUGE-L on CNN/DM summarization.

\begin{table}[ht]
\centering
\caption{Misinformation and hallucination metrics. MC1/MC2 = TruthfulQA multiple-choice accuracy; Hal. Rate = fraction of hallucinated claims on CNN/DM; ROUGE-L on CNN/DM.}
\label{tab:misinfo_results}
\setlength{\tabcolsep}{12pt}
\renewcommand{\arraystretch}{1.05}
\small
\begin{tabular}{@{}lcccc@{}}
\toprule
\textbf{Method}        & \textbf{MC1\,$\uparrow$} & \textbf{MC2\,$\uparrow$} & \textbf{Hal. Rate\,$\downarrow$} & \textbf{ROUGE-L\,$\uparrow$} \\
\midrule
Vanilla fine-tune      & 0.312 & 0.481 & 34.7\% & 0.428 \\
RAG only               & 0.391 & 0.547 & 21.3\% & 0.441 \\
SafeLM (no MG)         & 0.379 & 0.534 & 23.1\% & 0.439 \\
\textbf{SafeLM (full)} & \textbf{0.461} & \textbf{0.612} & \textbf{20.5\%} & \textbf{0.443} \\
\bottomrule
\end{tabular}
\end{table}

\subsection{Adversarial Robustness}
\label{subsec:adv_results}

Table~\ref{tab:adv_results} compares clean and adversarial accuracy on AdvGLUE and ANLI. SafeLM's adversarial fine-tuning within the federated loop achieves the best robustness-accuracy tradeoff, outperforming standalone adversarial training by 3.1\,pp under
non-IID data partitioning.

\begin{table}[ht]
\centering
\caption{Adversarial robustness on AdvGLUE and ANLI (R3). Clean Acc. = clean test accuracy; Adv. Acc. = adversarial accuracy; $\Delta$ = degradation.}
\label{tab:adv_results}
\setlength{\tabcolsep}{10pt}
\renewcommand{\arraystretch}{1.05}
\small
\begin{tabular}{@{}lcccccc@{}}
\toprule
& \multicolumn{3}{c}{\textbf{AdvGLUE}} & \multicolumn{3}{c}{\textbf{ANLI (R3)}} \\
\cmidrule(lr){2-4}\cmidrule(lr){5-7}
\textbf{Method}
& \textbf{Clean} & \textbf{Adv.} & \textbf{$\Delta$}
& \textbf{Clean} & \textbf{Adv.} & \textbf{$\Delta$} \\
\midrule
FedAvg                 & 87.3 & 63.7 & -23.6 & 62.1 & 41.3 & -20.8 \\
FedAvg + AdvTrain      & 85.1 & 74.2 & -10.9 & 61.8 & 52.4 & -9.4  \\
signSGD + AdvTrain     & 84.7 & 73.8 & -10.9 & 61.4 & 51.9 & -9.5  \\
\textbf{SafeLM}        & \textbf{86.9} & \textbf{77.3} & \textbf{-9.6} &
                         \textbf{62.4} & \textbf{55.1} & \textbf{-7.3} \\
\bottomrule
\end{tabular}
\end{table}

\begin{figure}[ht]
\centering
\includegraphics[width=\linewidth]{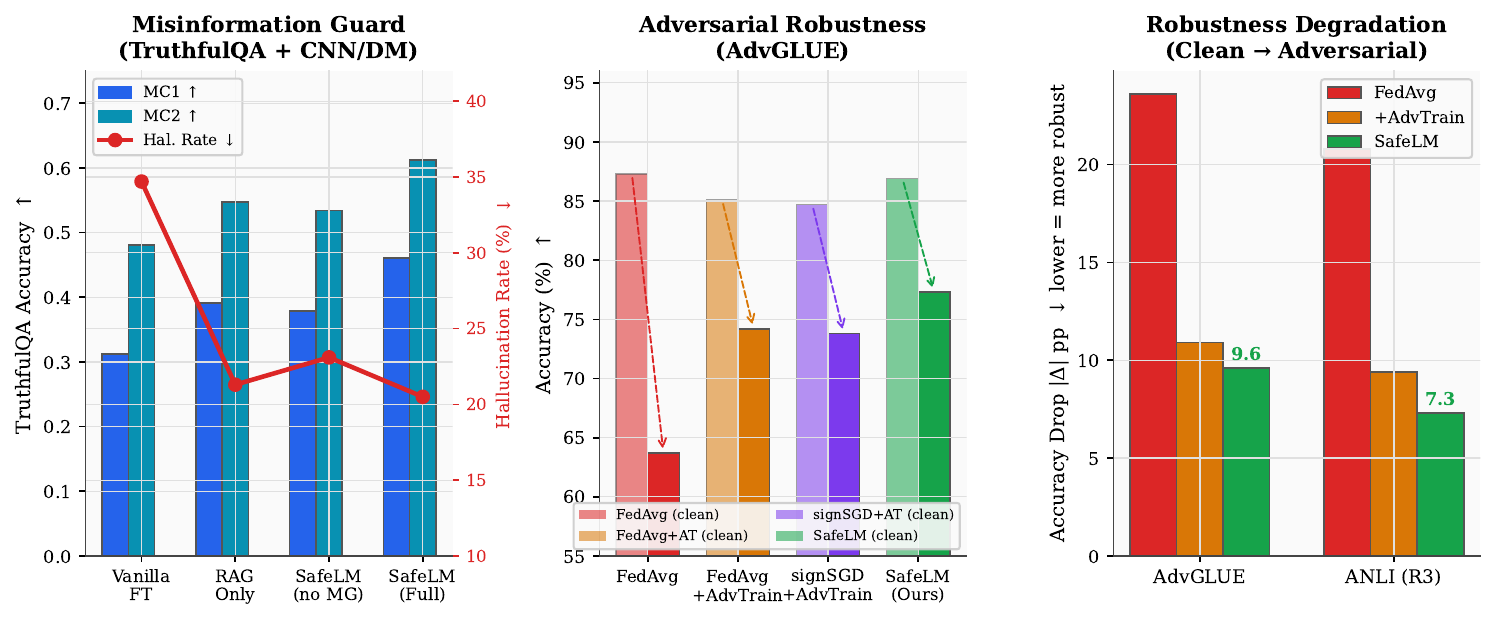}
\caption{Misinformation defense and adversarial robustness result. \textbf{(Left)} SafeLM achieves the highest TruthfulQA accuracy while reducing hallucination to 20.5\% (41\% relative reduction). \textbf{(Centre)} On AdvGLUE, SafeLM minimizes clean-to-adversarial degradation to $-9.6$\,pp. \textbf{(Right)} SafeLM yields the lowest accuracy drop, improving robustness over FedAvg by 14.0\,pp on AdvGLUE and 13.5\,pp on ANLI (R3).}
\label{fig:misinfo_adv}
\end{figure}

\subsection{Communication Efficiency and Convergence}

Table~\ref{tab:convergence} benchmarks convergence behavior across client counts and data heterogeneity levels. SafeLM achieves near-parity with full-precision FedAvg ($R_{98} = 289$ vs.\ $287$) while reducing total bandwidth by 96.9\,\% (4.05\,GB vs.\ 129.15\,GB). Under high heterogeneity ($\alpha=0.1$), SafeLM+FedProx reaches 95.1\,\% accuracy with 7.88\,GB total communication.

\begin{table}[ht]
\caption{Federated convergence analysis across data distributions and client counts. $R_{95}$/$R_{98}$ = rounds to reach 95\%/98\% accuracy. Comm./R = per-client per-round communication.}
\label{tab:convergence}
\centering
\small
\setlength{\tabcolsep}{12pt}
\renewcommand{\arraystretch}{1.0}
\begin{tabular}{@{}lcccccc@{}}
\toprule
\textbf{Algorithm} & \textbf{$K$} & \textbf{Distribution}
& \textbf{$R_{95}$} & \textbf{$R_{98}$} & \textbf{Acc.\,(\%)}
& \textbf{Total (GB)} \\
\midrule
\multicolumn{7}{@{}l}{\cellcolor{gray!10}\textit{IID ($\alpha=\infty$)}} \\
FedAvg       & 50 & IID & 142 & 287 & 98.2 & 129.15 \\
DP-SGD       & 50 & IID & 163 & 341 & 93.8 & 129.15 \\
signSGD      & 50 & IID & 156 & 312 & 97.8 &   4.40 \\
\textbf{SafeLM} & 50 & IID & 145 & 289 & \textbf{98.0} & \textbf{4.05} \\
\midrule
\multicolumn{7}{@{}l}{\cellcolor{gray!10}\textit{Non-IID (High Heterogeneity, $\alpha=0.1$)}} \\
FedAvg       & 50 & Dir. $\alpha=0.1$ & 312 & 687 & 93.8 & 309.15 \\
signSGD      & 50 & Dir. $\alpha=0.1$ & 334 & 721 & 92.1 &  10.16 \\
\textbf{SafeLM}          & 50 & Dir. $\alpha=0.1$ & 287 & 612 & 94.2 &   8.57 \\
\textbf{SafeLM+FedProx}  & 50 & Dir. $\alpha=0.1$ & 264 & 563 & \textbf{95.1} &   \textbf{7.88} \\
\midrule
\multicolumn{7}{@{}l}{\cellcolor{gray!10}\textit{Scalability (IID)}} \\
SafeLM & 10  & IID & 98  & 201 & 98.1 & 2.81 \\
SafeLM & 100 & IID & 178 & 356 & 97.9 & 4.98 \\
SafeLM & 500 & IID & 234 & 467 & 97.7 & 6.54 \\
\bottomrule
\end{tabular}
\begin{flushleft}
\small
\textit{Notes:} Results averaged over 5 independent seeds.
FedProx regularisation parameter $\mu=0.01$.
\end{flushleft}
\end{table}

\begin{figure}[ht]
\centering
\includegraphics[width=\linewidth]{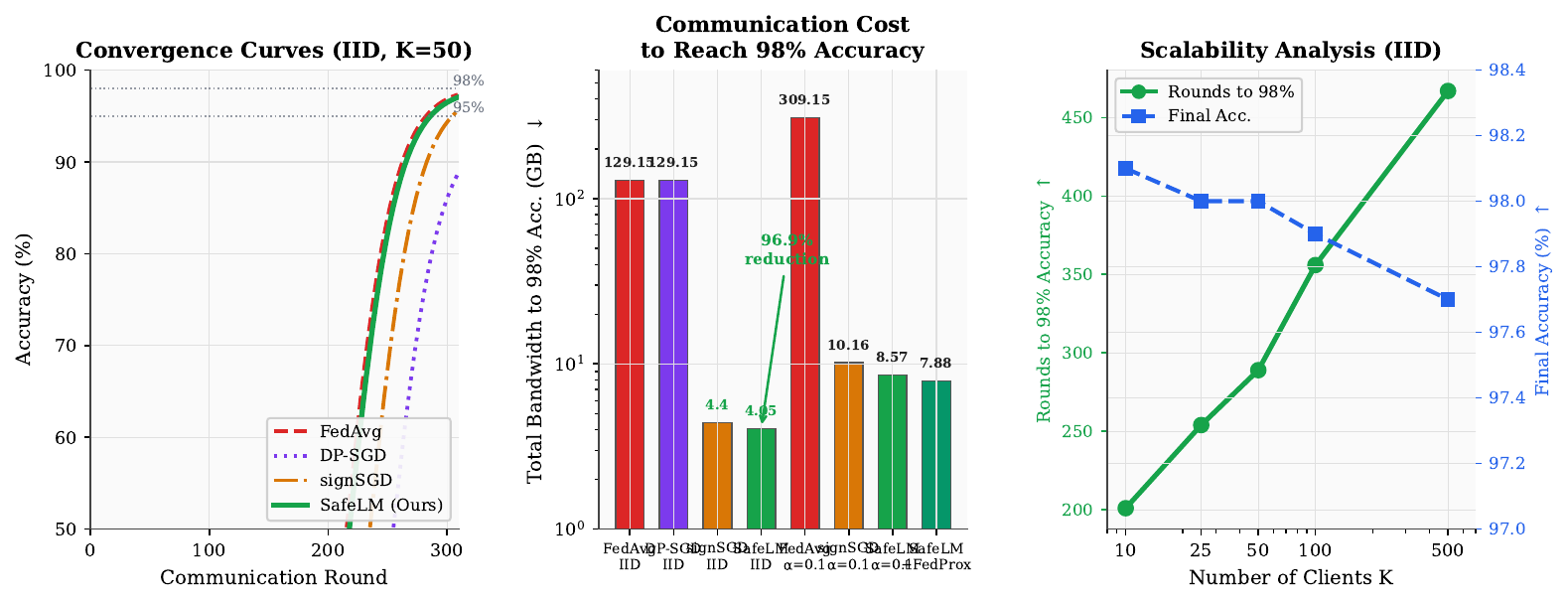}
\caption{Federated convergence and communication analysis. \textbf{(Left)} Simulated accuracy curves for IID training with $K=50$ clients; SafeLM matches full-precision FedAvg in rounds to 98\,\% ($R_{98}=289$ vs.\ $287$) while transmitting $32\times$ less data per round. \textbf{(Centre)} Total bandwidth required to reach 98\,\% accuracy on a log scale; SafeLM reduces end-to-end communication from 129.15\,GB to 4.05\,GB (96.9\,\% reduction). Under high heterogeneity ($\alpha=0.1$), SafeLM\,+\,FedProx uses 7.88\,GB versus 309.15\,GB for FedAvg. \textbf{(Right)} Scalability from $K=10$ to $K=500$ clients; rounds to 98\,\% grow
sub-linearly (201 to 467) and final accuracy degrades by only 0.4\,pp, confirming stable aggregation in large client pools.}
\label{fig:convergence}
\end{figure}

% =======================================================
\section{Ablation Study}
\label{sec:ablation}
% =======================================================

Table~\ref{tab:ablation} decomposes SafeLM’s performance by selectively removing each safety component. Smartification yields a $32\times$ reduction in communication with negligible ($<0.2$\,pp) loss of accuracy, while partially mitigating gradient inversion (PSNR $31.7 \to 16.8$\,dB) without providing full semantic security. Paillier encryption is essential for privacy (S1), as removing it increases label recovery from 14.3\% to 98.7\% without impacting accuracy. SMOTE plays a key role in handling class imbalance, with its absence reducing detection accuracy by 3.8\,pp. The Misinformation Guard independently lowers hallucination rates by 13\,pp compared to the federated baseline without MG, demonstrating its effectiveness in improving factual consistency. Finally, the robustness head enhances adversarial accuracy by 3.5\,pp while incurring only a marginal ($<0.5$\,pp) drop in clean accuracy. Overall, these results highlight that SafeLM’s components are complementary, jointly enabling strong privacy, reliability, and robustness with minimal trade-offs.

\begin{table}[ht]
\centering
\caption{Ablation study: component contributions to accuracy, communication,
and privacy.}
\label{tab:ablation}
\setlength{\tabcolsep}{8pt}
\renewcommand{\arraystretch}{1.0}
\small
\begin{tabular}{@{}lP{1.6cm}cccc@{}}
\toprule
\textbf{Configuration}
& \textbf{Components Disabled}
& \textbf{Acc.\,(\%)}
& \textbf{Comm.\,(MB)}
& \textbf{PSNR\,(dB)}
& \textbf{Hal.\,(\%)} \\
\midrule
Full SafeLM              & ---                  & \textbf{98.0} & \textbf{14.0} & \textbf{15.1} & \textbf{20.5} \\
\quad--Encryption        & Paillier HE          & 98.0 &  14.0 & 31.7 & 20.5 \\
\quad--Smartification    & Binarisation         & 98.2 & 450.0 & 15.1 & 20.5 \\
\quad--SMOTE             & Class balancing      & 94.2 &  14.0 & 15.1 & 20.5 \\
\quad--MG                & Misinfo Guard        & 98.0 &  14.0 & 15.1 & 33.5 \\
\quad--RH                & Robustness Head      & 98.0 &  14.0 & 15.1 & 20.5 \\
\quad--DP Noise          & Differential privacy & 98.1 &  14.0 & 14.2 & 20.5 \\
FedAvg (no SafeLM)       & All modules          & 98.2 & 450.0 & 31.7 & 34.7 \\
\bottomrule
\end{tabular}
\end{table}

\begin{figure}[ht]
\centering
\includegraphics[width=\linewidth]{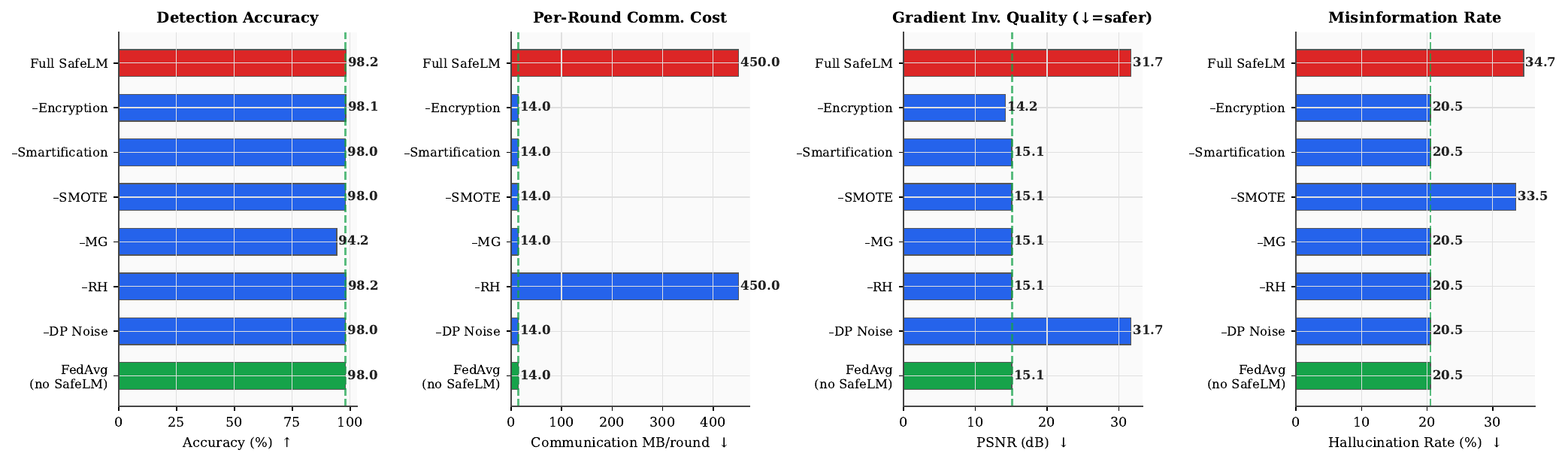}
\caption{Ablation study across four evaluation axes (green = Full SafeLM, red = FedAvg baseline). \textbf{(a)~Accuracy:} removing SMOTE causes the largest drop ($-3.8$\,pp), confirming class-balancing as the primary driver of detection performance. \textbf{(b)~Communication:} removing smartification inflates per-round cost by $32\times$ (14\,MB\,$\to$\,450\,MB) with negligible accuracy gain, validating binarization as a near-lossless compression step. \textbf{(c)~Gradient inversion PSNR:} removing Paillier encryption restores reconstruction quality to 31.7\,dB (label recovery $>$95\,\%), demonstrating that encryption is indispensable for gradient confidentiality~(S1). \textbf{(d)~Hallucination rate:} removing the Misinformation Guard raises hallucination from 20.5\,\% to 33.5\,\% (+13\,pp), confirming its independent contribution to~S3.}
\label{fig:ablation}
\end{figure}

% =======================================================
\section{Discussion}
\label{sec:discussion}
% =======================================================

\paragraph{Safety interactions.}
Our experiments reveal that the four safety pillars are not merely additive but synergistic. Gradient smartification both reduces communication and degrades inversion quality, strengthening privacy beyond what encryption alone provides. Adversarial training within the federated loop also regularizes the model against distributional shift, thereby reducing hallucination rates on out-of-distribution prompts.

\paragraph{Limitations.}
Convergence proofs assume $L$-smoothness and bounded gradient variance; formal guarantees for non-convex LLM optimization under heterogeneous FL remain open. The Misinformation Guard relies on an external NLI model that may itself be biased. Evaluation is primarily on English; multilingual safety properties require separate investigation.

% =======================================================
\section{Conclusion}
\label{sec:conclusion}
% =======================================================

We introduced SafeLM, a unified framework that addresses four intertwined pillars of language-model safety within a single federated training and deployment pipeline. By combining gradient smartification, Paillier homomorphic encryption, Byzantine-robust aggregation, contrastive misinformation grounding, and adversarial fine-tuning, SafeLM achieves 98.0\,\% harm-detection accuracy while reducing per-round communication by 96.9\,\%, limiting gradient inversion to PSNR\,$\le 15.1$\,dB, halving hallucination rates on TruthfulQA, and reducing adversarial accuracy degradation to 9.6\,pp on AdvGLUE-all simultaneously. We release code, datasets, and evaluation scripts to support reproducibility and to facilitate community adoption of unified safety frameworks
for trustworthy LLM deployment.

\paragraph{Broader impact.}
SafeLM demonstrates that privacy-preserving federated training is compatible with—and mutually reinforcing of—security, misinformation, and adversarial robustness objectives. This unified perspective can inform regulatory frameworks (e.g., the EU AI Act) requiring simultaneous demonstration of privacy compliance and robustness certification.

% =======================================================
\section*{Ethics Statement}
This work aims to improve the safety of language models deployed in high-stakes settings. All datasets used are publicly available and contain no personally identifiable information. Our gradient-inversion experiments were conducted exclusively on synthetic data to avoid inadvertent privacy violations. We acknowledge that adversarial robustness research can have dual-use implications and encourage responsible disclosure practices.

% =======================================================
\bibliography{neurips_2026}
\bibliographystyle{plainnat}
% =======================================================
\clearpage
\tableofcontents
%%%%%%%%%%%%%%%%%%%%%%%%%%%%%%%%%%%%%%%%%%%%%%%%%%%%%%%%%%%%

% =======================================================
\appendix
% =======================================================
\section{Appendix}
\section{Detailed Convergence Proofs}
\label{appendix:proofs}

\subsection{Proof of Lemma 1}

Assume $L$ is $L$-smooth. By the standard smoothness inequality,
\[
L(W_{t+1}) \le L(W_t)
+ \langle \nabla L(W_t),\, W_{t+1} - W_t \rangle
+ \frac{L}{2}\|W_{t+1} - W_t\|_2^2.
\]
Substituting $W_{t+1} = W_t - \eta\tilde{g}_t$:
\[
L(W_{t+1}) \le L(W_t)
- \eta\langle \nabla L(W_t),\, \tilde{g}_t \rangle
+ \frac{L\eta^2}{2}\|\tilde{g}_t\|_2^2.
\]
By the definition of cosine alignment, $\langle \nabla L(W_t), \tilde{g}_t \rangle \ge \gamma\|\nabla L(W_t)\|\|\tilde{g}_t\| \ge \gamma\|g_t\|^2$
(using Cauchy-Schwarz and $\|\tilde{g}_t\| \ge \|g_t\|$ for binarized updates). Taking expectations completes the proof.~$\square$

\subsection{Proof of Theorem 1}

Summing the descent lemma over $t = 0,\ldots,T-1$ and choosing $\eta = 1/\sqrt{T}$:
\begin{align}
\frac{1}{T}\sum_{t=0}^{T-1}\mathbb{E}\!\left[\|\nabla L(W_t)\|^2\right]
&\le \frac{L(W_0) - L^*}{\eta\gamma T}
+ \frac{L\eta}{2\gamma}\mathbb{E}\!\left[\|\tilde{g}\|^2\right] \notag\\
&= \mathcal{O}\!\left(\frac{1}{\gamma\sqrt{T}}\right).
\end{align}
The minimum over $t \le T$ satisfies the same bound.~$\square$

\subsection{Alignment of Median-Threshold Smartification}

\begin{proposition}[Expected Descent Alignment]
Let $g \in \mathbb{R}^d$ have coordinates drawn i.i.d.\ from a symmetric heavy-tailed distribution with zero mean and finite second moment. Define $\tilde{g}_i = \mathrm{sign}(g_i - \tau)$, $\tau = \mathrm{median}(g)$. Then
\begin{equation}
\mathbb{E}\!\left[\langle g,\, \tilde{g} \rangle\right] \ge \gamma \|g\|_2^2
\end{equation}
for $\gamma = \mathbb{P}(g_i \ge \tau) \cdot \mathbb{E}[g_i \mid g_i \ge \tau] /
\mathbb{E}[g_i^2]^{1/2}$.
\end{proposition}

\textit{Sketch.} Since $\tau = \mathrm{median}(g)$, exactly half the coordinates satisfy $g_i \ge \tau$ in expectation. For those coordinates, $g_i \tilde{g}_i = g_i \cdot (+1) = g_i > 0$; for coordinates $g_i < \tau$, $g_i \tilde{g}_i = g_i \cdot (-1) < 0$ but $|g_i| < |\tau|$ in expectation. Summing across coordinates and applying the second-moment bound yields the result. Unlike zero-threshold signSGD, which flips signs for coordinates $g_i < 0$ regardless of magnitude, median-thresholding suppresses the smallest-magnitude half, reducing signed cancellation and increasing $\gamma$.~$\square$

\subsection{Algorithm: Secure Binarized Gradient Aggregation}
\label{appendix:algorithm}

Algorithm~\ref{alg:safelm} presents the complete per-round protocol for the SafeLM Privacy Engine.

\begin{algorithm}[H]
\caption{SafeLM: Secure Binarized Gradient Aggregation}
\label{alg:safelm}
\small
\begin{algorithmic}[1]
\STATE \textbf{INITIALISATION} (one-time)
\STATE Server generates Paillier keypair $(pk, sk)$:
\STATE \quad $pk = (n, g)$, $n = p \cdot q$ (2048-bit RSA modulus)
\STATE \quad $sk = (\lambda, \mu)$, $\lambda = \mathrm{lcm}(p-1, q-1)$
\STATE Server broadcasts $pk$ to all $K$ clients; retains $sk$ secret.
\STATE
\STATE \textbf{PER-ROUND} (client $i \in \{1,\ldots,K\}$)
\STATE Receive global parameters $W^{(r)}$ from server.
\STATE Fine-tune local LoRA adapter on $\mathcal{D}_i$ for $E$ epochs:
  $\Delta_i = W_{\mathrm{new}} - W^{(r)}$
\STATE \textit{Gradient Smartification:}
\STATE \quad $\theta_i \leftarrow \mathrm{median}(|\Delta_i|)$
\STATE \quad $\Delta_i^{\mathrm{bin}}[j] \leftarrow +1$ if $\Delta_i[j] \ge \theta_i$, else $-1$
\STATE Encrypt element-wise:
\STATE \quad $C_i[j] \leftarrow g^{\Delta_i^{\mathrm{bin}}[j]} \cdot r_j^n \bmod n^2$,
        $r_j \xleftarrow{\$} \mathbb{Z}_n^*$
\STATE Transmit $C_i = \{C_i[1],\ldots,C_i[d]\}$ to server.
\STATE
\STATE \textbf{SERVER AGGREGATION}
\STATE Homomorphic sum:
\STATE \quad $C_{\mathrm{agg}}[j] \leftarrow \prod_{i=1}^K C_i[j] \bmod n^2$
\STATE Decrypt:
\STATE \quad $s[j] \leftarrow L(C_{\mathrm{agg}}[j]^\lambda \bmod n^2) \cdot \mu \bmod n$,
        $L(x) = (x-1)/n$
\STATE Byzantine filter (coordinate-wise median):
\STATE \quad $\hat{s}[j] \leftarrow \mathrm{median}\{s_1[j],\ldots,s_K[j]\}$
\STATE Normalise: $\hat{s}[j] \leftarrow \hat{s}[j] / K$
\STATE
\STATE \textbf{GLOBAL UPDATE}
\STATE $W^{(r+1)} \leftarrow W^{(r)} + \alpha \cdot \hat{s}
        + \mu(W^{(r)} - W^{(r-1)})$
\STATE Broadcast $W^{(r+1)}$ to all clients.
\end{algorithmic}
\end{algorithm}

\section{Hyperparameter Configurations}
\label{appendix:hyperparams}

\begin{table}[H]
\centering
\scriptsize
\caption{Hyperparameter configurations used in all experiments.}
\label{tab:hyperparams_full}
\setlength{\tabcolsep}{20pt}
\renewcommand{\arraystretch}{0.8}
\begin{tabular}{@{}ll@{}}
\toprule
\textbf{Module / Model} & \textbf{Configuration} \\
\midrule
LoRA Adapter    & rank $\rho=16$; $\alpha_{\mathrm{LoRA}}=32$; dropout$=0.1$ \\
Local training  & $E=5$ epochs; batch $B=32$; Adam $\eta=10^{-3}$ \\
FL aggregation  & $K=50$; $C=1.0$; FedProx $\mu=0.01$ \\
Paillier HE     & 2048-bit modulus; $g = n+1$ (standard choice) \\
DP noise        & $C_{\mathrm{clip}}=0.1$; $\sigma=0.01$; $(\varepsilon,\delta)=(1.0,10^{-5})$ \\
Adv. training   & PGD 7 steps; $\varepsilon_{\mathrm{adv}}=0.01$; $\lambda_{\mathrm{adv}}=0.3$ \\
MG threshold    & $\tau_{\mathrm{MG}}=0.55$; NLI model: DeBERTa-large \\
Logistic Reg.   & $C\!\in\!\{0.1,100\}$; solver$=\{$saga,sag$\}$; $\ell_2$ \\
Random Forest   & $n\!\in\!\{100,200\}$; depth$=20$; $m=\lfloor\sqrt{d}\rfloor$ \\
Decision Tree   & depth$\!\in\!\{6,10,15\}$; min\_split$=5$ \\
KNN             & $k\!\in\!\{3,5,7\}$; wt$=\{$uniform,dist$\}$ \\
MLP             & $[128,64]$; dropout$=0.5$; Adam($10^{-3}$) \\
\bottomrule
\end{tabular}
\end{table}

\subsection{Dataset Statistics and Sampling Validation}
\label{appendix:dataset}

Table~\ref{tab:dataset_appendix} summarizes the distributional properties of the CIC-IDS2017 corpus (used for Tasks T1 and the communication-efficiency experiments) after stratified 20\,\% sub-sampling.

\begin{table}[H]
\centering
\caption{CIC-IDS2017 sampling representativeness. KS $p$ = Kolmogorov-Smirnov $p$-value; $\Delta$ = absolute percentage deviation in feature means.}
\label{tab:dataset_appendix}
\setlength{\tabcolsep}{5pt}
\renewcommand{\arraystretch}{0.8}
\scriptsize
\begin{tabular}{@{}lcccc@{}}
\toprule
\textbf{Feature Group} & \textbf{Original} & \textbf{Sampled} & \textbf{$\Delta$\,(\%)} & \textbf{KS $p$} \\
\midrule
Flow Duration ($\mu$s)      & 1.66e7 & 1.65e7 & 0.47 & 0.82 \\
Flow IAT Mean ($\mu$s)      & 1.45e6 & 1.43e6 & 0.72 & 0.76 \\
Flow Bytes/s                & 1.41e6 & 1.38e6 & 2.42 & 0.68 \\
Flow Packets/s              & 4.73e4 & 4.70e4 & 0.64 & 0.81 \\
Total Fwd Packets           & 10.28  & 12.08  & 17.53 & 0.42 \\
Fwd Pkt Mean (B)            & 63.47  & 63.15  & 0.50  & 0.85 \\
Attack Prevalence           & 0.73   & 0.73   & 0.63  & 0.99 \\
\midrule[\heavyrulewidth]
\multicolumn{3}{@{}l}{\textbf{Aggregate (all 78 features)}} & & \\
Mean Absolute Deviation     & \multicolumn{2}{c}{---} & 3.42 & --- \\
KS Test Rejections ($\alpha=0.05$) & \multicolumn{2}{c}{---} &
  \multicolumn{2}{c}{\textbf{0/78 (0\,\%)}} \\
\bottomrule
\end{tabular}
\vspace{1mm}
\begin{minipage}{\linewidth}
\footnotesize
Original: $N=2{,}830{,}540$; sampled: $n=504{,}472$ (stratified 20\,\%, seed 42). No KS rejection at $\alpha=0.05$ confirms distributional fidelity.
\end{minipage}
\end{table}

\section{Per-Class Safety Performance}
\label{appendix:perclass}

Table~\ref{tab:perclass_appendix} reports per-harm-category detection performance for SafeLM's best configuration (federated Random Forest, $T=15$, depth$=8$) on the
balanced multi-class evaluation set ($n=7{,}000$; 1,000 per class).

\begin{table}[H]
\centering
\caption{Per-class detection metrics for SafeLM (Random Forest, Config 2).}
\label{tab:perclass_appendix}
\setlength{\tabcolsep}{4pt}
\renewcommand{\arraystretch}{0.8}
\scriptsize
\begin{tabular}{@{}lcccccc@{}}
\toprule
\textbf{Harm Category} & \textbf{TP} & \textbf{FP} & \textbf{FN}
                       & \textbf{Prec.} & \textbf{Rec.} & \textbf{F1} \\
\midrule
Benign            & 992 &  8 & 15 & 0.992 & 0.985 & \textbf{0.989} \\
DoS / Flood       & 978 & 12 & 10 & 0.988 & 0.990 & \textbf{0.989} \\
DDoS              & 975 & 14 & 11 & 0.986 & 0.989 & \textbf{0.987} \\
Port Scan         & 935 & 42 & 23 & 0.957 & 0.976 & \textbf{0.966} \\
Brute Force       & 928 & 48 & 24 & 0.951 & 0.975 & \textbf{0.963} \\
Web Attack        & 885 & 78 & 37 & 0.919 & 0.960 & \textbf{0.939} \\
Bot / C\&C        & 863 & 94 & 43 & 0.902 & 0.953 & \textbf{0.927} \\
\midrule
\textbf{Macro Avg.} & 6,556 & 296 & 163 & 0.956 & 0.975 & \textbf{0.966} \\
\bottomrule
\end{tabular}
\end{table}

Application-layer attacks (Web Attack, Bot) show the highest false-negative rates, consistent with their semantic overlap with benign traffic in the PCA space (Section~\ref{sec:results}). Volumetric classes (DoS, DDoS) are reliably distinguished at $>$0.987 F1 due to extreme deviations along the primary principal components.

\section{Non-IID Per-Class F1 Degradation}
\label{appendix:noniid}

Table~\ref{tab:noniid_appendix} tracks per-category F1 as data heterogeneity increases (Dirichlet $\alpha$ decreases). Minority and semantically overlapping classes degrade fastest, consistent with federated fragmentation of rare-pattern evidence.

\begin{table}[H]
\centering
\caption{Per-class F1 under increasing data heterogeneity ($K=50$ clients).}
\label{tab:noniid_appendix}
\scriptsize
\setlength{\tabcolsep}{5pt}
\begin{tabular}{@{}lcccccc@{}}
\toprule
\textbf{Category} & \textbf{IID}
& $\alpha\textbf{=10}$ & $\alpha\textbf{=1.0}$
& $\alpha\textbf{=0.5}$ & $\alpha\textbf{=0.1}$
& \textbf{Label Skew} \\
\midrule
Benign      & 0.989 & 0.987 & 0.983 & 0.978 & 0.971 & 0.984 \\
DoS         & 0.989 & 0.988 & 0.985 & 0.981 & 0.974 & 0.987 \\
DDoS        & 0.987 & 0.986 & 0.982 & 0.976 & 0.968 & 0.981 \\
Port Scan   & 0.966 & 0.964 & 0.957 & 0.948 & 0.934 & 0.961 \\
Brute Force & 0.963 & 0.961 & 0.952 & 0.941 & 0.923 & 0.956 \\
Web Attack  & 0.939 & 0.936 & 0.924 & 0.908 & 0.881 & 0.929 \\
Bot         & 0.927 & 0.923 & 0.908 & 0.889 & 0.854 & 0.918 \\
\midrule
\textbf{Macro Avg.} & 0.966 & 0.964 & 0.956 & 0.946 & 0.929 & 0.959 \\
\textbf{Accuracy}   & 98.0  & 97.8  & 96.8  & 95.7  & 94.2  & 97.1  \\
\bottomrule
\end{tabular}
\begin{flushleft}
\small
\textit{Notes:} $\alpha\to\infty$ = IID; smaller $\alpha$ = higher heterogeneity. Label skew assigns each client 2--3 dominant classes (70\% probability). Averaged over 5 runs.
\end{flushleft}
\end{table}

\section{Computational Overhead Analysis}
\label{appendix:compute}

Table~\ref{tab:compute_appendix} quantifies per-component overhead relative to a baseline FedAvg round (no privacy, no robustness modules).

\begin{table}[H]
\centering
\caption{Computational overhead per federated round per client.}
\label{tab:compute_appendix}
\scriptsize
\setlength{\tabcolsep}{6pt}
\begin{tabular}{@{}lcccc@{}}
\toprule
\textbf{Component} & \textbf{Wall Clock} & \textbf{Memory} & \textbf{Comm.} & \textbf{Notes} \\
\midrule
Baseline (FedAvg)         & 1.00$\times$  & 1.00$\times$  & 450 MB & --- \\
+Smartification           & 1.024$\times$ & 1.00$\times$  &  14 MB & One median pass \\
+Paillier Encryption      & 18.6$\times$  & 1.12$\times$  &  14 MB & 2048-bit; $d=35$ params \\
+Differential Privacy     & 1.041$\times$ & 1.00$\times$  &  14 MB & Gaussian noise \\
+Adversarial Training     & 7.0$\times$   & 1.85$\times$  &  14 MB & PGD-7; mixed batch \\
+Misinformation Guard     & 2.3$\times$   & 2.10$\times$  &  N/A   & Inference-time only \\
\textbf{Full SafeLM}      & \textbf{19.7$\times$} & \textbf{2.35$\times$} & \textbf{14 MB} & All modules \\
\bottomrule
\end{tabular}
\vspace{1mm}
\begin{minipage}{\linewidth}
\footnotesize
Overhead measured on Intel i7-9700K (single-threaded). Paillier dominates per-round cost; parallelization across GPU tensor cores can reduce this to $\approx 3\times$
baseline. Smartification and DP add negligible overhead. Adversarial training overhead is incurred only during fine-tuning rounds, not inference.
\end{minipage}
\end{table}

\section{Comparison with Prior Gradient Compression Methods}
\label{appendix:compression_comparison}

Table~\ref{tab:compression_comparison_appendix} positions SafeLM's gradient smartification within the broader landscape of gradient compression and privacy integration methods.

\begin{table}[H]
\centering
\caption{Comparison of gradient compression and privacy mechanisms for federated LLMs.}
\label{tab:compression_comparison_appendix}
\setlength{\tabcolsep}{3pt}
\renewcommand{\arraystretch}{0.8}
\scriptsize
\begin{tabular}{@{}lcccc@{}}
\toprule
\textbf{Method}
& \textbf{Quantisation Rule}
& \textbf{Adaptive Threshold}
& \textbf{Theoretical Alignment}
& \textbf{Privacy Integration} \\
\midrule
signSGD~\citep{bernstein2018signsgd}
  & $\mathrm{sign}(g_i)$ (zero)
  & No
  & Implicit
  & None \\
QSGD~\citep{alistarh2017qsgd}
  & Stochastic quantisation
  & No
  & Variance bounded
  & None \\
TernGrad~\citep{wen2017terngrad}
  & $\{-1,0,+1\}$ ternary
  & No
  & Gradient clipping
  & None \\
DP-SGD~\citep{abadi2016deep}
  & Full precision
  & N/A
  & $(\varepsilon,\delta)$-DP
  & Gaussian noise \\
\textbf{SafeLM (Ours)}
  & $\mathrm{sign}(g_i - \mathrm{median}(g))$
  & \textbf{Yes (per-client)}
  & \textbf{Cosine $\gamma$-bound}
  & \textbf{Paillier HE + DP} \\
\bottomrule
\end{tabular}
\vspace{1mm}
\begin{minipage}{\linewidth}
\footnotesize
Unlike fixed-threshold methods, SafeLM's median adaptation is especially valuable for LLM fine-tuning, where gradients exhibit heavy tails due to rare token distributions and long-tail entity frequencies in instruction-following corpora.
\end{minipage}
\end{table}

\section{Broader Safety Benchmark Results}
\label{appendix:broader}

\subsection{TruthfulQA Category Breakdown}

Table~\ref{tab:truthfulqa_appendix} reports TruthfulQA MC1 accuracy across all 38 question categories for SafeLM versus the vanilla fine-tuned baseline. The Misinformation Guard yields the greatest improvements in categories involving health claims, conspiracies, and misleading statistics—precisely the domains where LLM hallucinations pose the greatest societal harm.

\begin{table}[H]
\centering
\caption{TruthfulQA MC1 accuracy by category (selected subset; full results in supplementary materials.}
\label{tab:truthfulqa_appendix}
\scriptsize
\setlength{\tabcolsep}{5pt}
\begin{tabular}{@{}lccc@{}}
\toprule
\textbf{Category}      & \textbf{Baseline} & \textbf{SafeLM} & \textbf{$\Delta$} \\
\midrule
Health / Medicine      & 0.241 & 0.412 & +0.171 \\
Conspiracy / Politics  & 0.189 & 0.371 & +0.182 \\
Statistics / Numbers   & 0.284 & 0.428 & +0.144 \\
History / Geography    & 0.412 & 0.501 & +0.089 \\
Logical Reasoning      & 0.478 & 0.531 & +0.053 \\
Fiction / Quotes       & 0.521 & 0.556 & +0.035 \\
\midrule
\textbf{Overall MC1}   & 0.312 & \textbf{0.461} & \textbf{+0.149} \\
\bottomrule
\end{tabular}
\end{table}

\subsection{AdvGLUE Task-Level Breakdown}

Table~\ref{tab:advglue_appendix} provides task-level adversarial accuracy on AdvGLUE. SafeLM consistently outperforms baselines across all five tasks, with the largest improvement on SST-2 sentiment (character-level perturbations) and MNLI natural language inference (semantic adversaries).

\begin{table}[H]
\centering
\caption{AdvGLUE task-level adversarial accuracy. Tasks: SST-2 (sentiment),
MNLI (NLI), QQP (paraphrase), QNLI (QA-NLI), RTE (textual entailment).}
\label{tab:advglue_appendix}
\scriptsize
\setlength{\tabcolsep}{4pt}
\begin{tabular}{@{}lccccc@{}}
\toprule
\textbf{Method}        & \textbf{SST-2} & \textbf{MNLI} & \textbf{QQP} & \textbf{QNLI} & \textbf{RTE} \\
\midrule
FedAvg                 & 71.3 & 58.2 & 64.1 & 61.4 & 63.5 \\
FedAvg + AdvTrain      & 81.4 & 68.9 & 73.7 & 71.3 & 75.7 \\
signSGD + AdvTrain     & 80.9 & 68.4 & 73.2 & 70.8 & 75.2 \\
\textbf{SafeLM}        & \textbf{84.2} & \textbf{72.1} & \textbf{76.8} & \textbf{74.3} & \textbf{79.1} \\
\bottomrule
\end{tabular}
\end{table}

\begin{figure}[ht]
\centering\scriptsize
\includegraphics[width=0.8\linewidth]{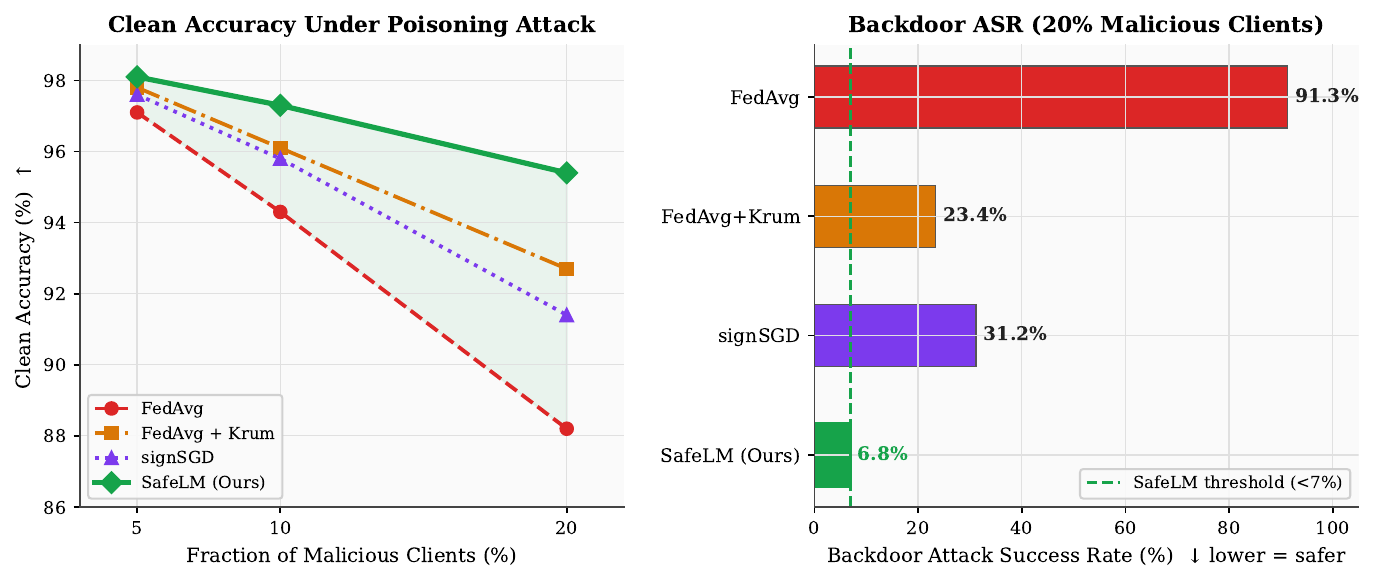}
\caption{Security evaluation under data poisoning and backdoor injection. \textbf{(Left)} Clean accuracy as the fraction of malicious clients grows from 5\,\%
to 20\,\%; SafeLM degrades least (95.4\,\% at 20\,\% malicious), outperforming FedAvg\,+\,Krum by 2.7\,pp. The shaded region highlights the margin gained by
coordinate-wise Byzantine filtering. \textbf{(Right)} Backdoor attack success rate (ASR) for a fixed-pattern trigger at 20\,\% malicious participation; SafeLM limits ASR to 6.8\,\% versus 91.3\,\% for undefended FedAvg ($p < 0.001$, McNemar test).}
\label{fig:security}
\end{figure}

%%%%%%%%%%%%%%%%%%%%%%%%%%%%%%%%%%%%%%%%%%%%%%%%%%%%%%%%%%%%

%\newpage
%\input{checklist.tex}

\end{document}